# Integrated Photonic Accelerator Based on Optical Spectrum Slicing for Convolutional Neural Networks

Aris Tsirigotis,[1] George Sarantoglou,[1] Stavros Deligiannidis,[2] Kostas Sozos,[2] Adonis Bogris,[2] and Charis Mesaritakis[1,*]

[1] Dept. Information and Communication Systems Engineering, University of the Aegean, Palama 2 str. Karlovasi Samos 83200-Greece
[2] Dept. Informatics and Computer Engineering, University of West Attica, Ag. Spiridonos, 12243, Egaleo, Greece

*Corresponding author: cmesar@aegean.gr

**In this work we numerically analyze a passive photonic integrated neuromorphic accelerator based on hardware-friendly optical spectrum slicing nodes. The proposed scheme can act as a fully analogue convolutional layer, preprocessing information directly in the optical domain. The proposed scheme allows the extraction of meaningful spatio-temporal features from the incoming data, thus when used prior to a simple fully connected digital single layer network it can boost performance with negligible power consumption. Numerical simulations using the MNIST dataset confirmed the acceleration properties of the proposed scheme, where 10 neuromorphic nodes can replace the convolutional layers of a sophisticated LeNet-5 network, thus reducing the number of total floating point operations per second (FLOPS) by 98% while offering a 97.2% classification accuracy.  ©2022 The Author(s)**

The exploding growth of the Internet of everything (IoE) ecosystem [1] has unleashed the generation of a tremendous amount of raw data that need processing so as to extract meaningful information. In this landscape, typical von-Neumann machines have met an efficiency road-block [2] and bio-inspired computing has emerged as an unconventional route aiming to circumvent inherent limitations. In this context, silicon photonics can offer a proliferating platform, based on merits such as wavelength/time multiplexing assisted parallelism, marginal power consumption, zero latency and the alleviation of bandwidth/fan-out/in trade off that plague electronic neuromorphic schemes [3–5].

Until recently, photonic neuromorphic schemes were limited to simple tasks or they were targeting niche but narrow-scope applications, where the target signals were inherently optical (optical equalization e.g.) [6,7]. Taking into consideration that the vast majority of raw data are in a digital format, a critical question is how photonics can infiltrate this area. Towards this direction, photonic computing schemes have currently harnessed considerable momentum in complex, time sensitive applications on cloud and fog/edge computing [8–10]; due to their computing efficiency that scales to the femtojoule per multiply-and-accumulate operation (MAC) [11]. Furthermore, the most promising role for photonic neuromorphic schemes is acceleration; meaning that the photonic circuit is restricted only to a demanding part of the computation, instead of trying to replace the full-scale digital system. Through this approach, the open issue of neural training in the optical domain is circumvented, whereas compliance to the merits of both worlds (analogue and digital) is achieved.

Based on the fact that one of the most common types of IoE data are images, convolutional neural networks (CNNs) that have been inspired by the operation of the visual cortex neurons to detect light in receptive fields [12], have risen as the most effective candidate for image recognition and processing. In general, CNN architectures contain three types of layers: convolutional, pooling and fully-connected layers. At the convolutional layers, dot product operations are performed between the input tensor values and matrices of different weights, namely kernels, whereas a non-linear activation function (e.g. ReLU) is applied to these elements. As a result, multiple "feature maps" are presented at the output of each layer. Among these layers' operations, convolution is the most computational "expensive". That is the reason why traditional graphics processing units (GPUs) [13], memristors [14] and lately photonic implementations focus on optimizing/replacing standard digital matrix-to-vector multiplications (MVMs). According to this approach, the image/feature map values are divided in patches which are serialized and arranged in the rows of a matrix, whereas kernel values are concentrated in the columns of another matrix with each kernel occupying a different column. Integrated photonic schemes implementing MVMs are based on microring resonator (MRR) banks [15,16], photonic tensor cores exploiting frequency-comb encoded input or kernel weights [17,18] and cascaded Mach Zehnder interferometers (MZI) meshes [3,19]. In all these works, the values for the MVM are computed/trained with the help of cumbersome physically accurate numerical simulations and off-line training.

In this work, we present an unconventional photonic accelerator, suitable for CNNs, that relies on optical spectral slicing (OSS) to emulate a convolutional process in the analog domain using complex weights. Similar filter based preprocessing has been demonstrated in the past for telecom applications; either followed by a digital reservoir computing (RC) [20] or as in [21] where an all-optical recurrent OSS was used alongside digital linear regression. The accelerator in this work consists of multiple parallel bandpass optical filters, where each can extract different spectro-temporal features from the input tensors (images), offering signal diversity. More importantly, the complex weights that are applied are not needed to be fine-tuned, as in the case of previous accelerators, but are only subject to course adjustments, similar to RC's hyperparameters.

As depicted in Fig.1, incoming information is imprinted to the amplitude of an optical carrier, provided by a conventional coherent source, through a typical Mach-Zehnder modulator. The back-end of the architecture consists of multiple, passive OSS nodes that are followed by a photodiode (PD) and an analogue-to-digital converter (ADC). The filter-based nodes' central frequency and bandwidth are set so as to target at different spectral regions of the input, acting in this way as different CNN kernels. The bandwidth and the central frequencies are set as to fully cover the bandwidth of the signal and can also vary according to the number of employed nodes ($N$). A PD is placed after each node, followed by an ADC. The digital outputs are flattened and feed a typical digital FCL and softmax layer. It should be noted here that typical training algorithms, such as backpropagation, need only to be applied at the digital layers, whereas the photonic OSS parameters, such as filter bandwidth, order and central frequency can be treated as trainable hyperparameters.

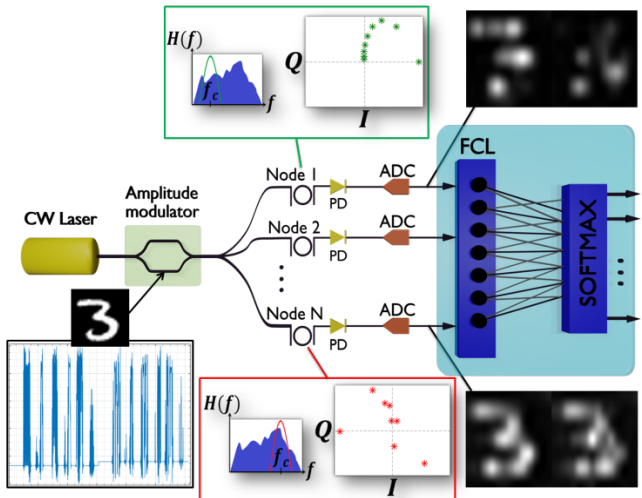

Fig. 1. Schematic diagram of the OSS-CNN architecture. At the insets the different spatial features for the digit "3" are presented by changing the central position of the filter-nodes.

In order to validate the efficiency of the proposed scheme as CNN accelerator we target an image classification task based on the MNIST dataset [22]. The MNIST images (60000 for training and 10000 for testing) are not digitally preprocessed and are only subject to a simple pixel rearranging patching scheme. More specifically, each image is divided into non-overlapping patches of the same size ($nxn$) which are serialized with two orientations (Fig.2), one column-wise (A) and one row-wise (B). Both of these orientations are employed from each patch and merged together as to produce a single vector for each image.

The OSS filters can be implemented with the use of MRRs, MZIs, complex setups [21] or even with generic first-order filters. In this work, we utilize such a generic approach, which is similar to first order MRR filters, whose drop port is employed as nodal output.

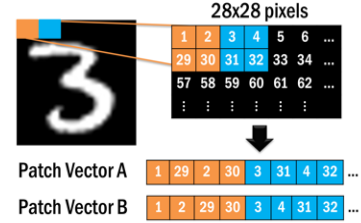

Fig. 2. 2x2 patching of the MNIST images and patch serialization with two orientations: A and B

In general, the time impulse response of a first-order filter is given by eq. 1, where $f_m$ and $f_c$ correspond to the central and the cut-off frequency of the bandpass filter respectively.

$$h(t) = 2\pi f_c e^{-t(2\pi f_c - j2\pi f_m)}, \ t > 0 \quad (1)$$

As it can be seen, filters with a different frequency detuning from the optical carrier (and/or a different bandwidth) offer diversified impulse responses (see Fig.1), corresponding to convolutions with different complex values between the filter-nodes and the signal. Accordingly, the receptive field; meaning the spatio-temporal window upon CNNs operate, can be regulated by the bandwidth and the order of the nodal filters. In particular, wider bandwidth filters induce a faster exponential decay of the impulse response and therefore a smaller number of pixels will participate at each convolutional operation, while the filter's order defines the steepness of the filters and thus the asymptotic decay of the applied weights within the time interval of interest.

The PDs after each filter through the square law offer a ReLU like non-linear activation function on each "feature map" [23].Taking into account the effects of shot and thermal noise, the PDs are also simulated to be followed by fourth-order Butterworth filters of a 3dB bandwidth which is inversely proportional to the employed patch size and is given by:

$$BW_{PD} = \frac{PR}{n^2} \quad (2)$$

where $PR$ is the pixel rate of the input signal and $n^2$ is the number of total patch elements (patch size). In this manner, an average value that corresponds to a timeslot of $n^2$ initial pixels is produced at the output of each PD as the optical signal passes through. Therefore, a data averaging process is carried out at the PDs, similar to the average pooling operation of a typical CNN. Following the PDs, an 8-bit precision ADC is assumed with a sampling rate ($SR$) that is initially set at two samples per patch ($SR=2BW_{PD}$), which is the Nyquist limit dictated by the bandwidth of the PDs. The digitized signals from each node are flattened and fed to the digital FCL, whereas the total number of samples defines the number of synapses (connections) for the FCL. The last stage consists of performing offline training at the FCL using conventional backpropagation with an Adam optimizer.

The effect of patching scheme size ($nxn$), the number of OSS nodes $N$ and the OSS filter characteristics $f_c$, $f_m$ were investigated along with the total input power and the sampling rate ($SR$) of the

ADCs so as to identify under which conditions classification accuracy, parameter minimization, throughput and power consumption can be optimized. The bandwidth and central frequencies of the nodal bandpass filters are properly selected to fully cover the input's signal spectrum with minimal overlap depending on the number of OSS-CNN nodes, which were set to $N = 2, 3, 5$ and 10. Assuming a state-of-the-art 128 GSa/s DAC and a 5 node OSS-CNN, the filters cut-off and central frequencies are respectively set to $f_c = 6.4$ GHz and $f_m = 6.4, 19.2, 32, 44.8, 57.6$ GHz. In fig. 3, classification accuracy is presented as a function of the compression ratio for different number of filter nodes (fig. 3a) and for different patch sizes (fig. 3b). Compression ratio defined as the ratio of the size of the input tensor (28x28 for the MNIST images) to the FCL's input size (flattened image sequences size) is a useful tool to demonstrate the parameter reduction that is carried out by the OSS-CNN compared to a standalone FCL. The sampling rate of the ADCs is also varied starting from the Nyquist limit down to an under-sampling regime, regulating in this way the dimensionality reduction degree which translates to a different number of samples at the input of the FCL and a different compression ratio. A standalone FCL fed directly with all the MNIST image pixels yields an accuracy of 92.1% (see Fig.3). Using OSS before the FCL leads to an improvement of classification with the number of employed nodes, reaching a maximum of 97.6% with 10 OSS nodes, whereas using more nodes provided no improvement in terms of accuracy (see Fig.3a). The size of the patching scheme ($nxn$) is also explored for $n = 2, 3, 4, 5$ with the 4x4 patch serialization delivering eventually the maximum accuracy for the MNIST images in all node cases. Most importantly, OSS-CNN was able to achieve an accuracy of 94.2%, 96.1% and 97.6% with a compression ratio of 4, 2 and 0.8 respectively compared to a standalone FCL (92.1% accuracy), proving that OSS facilitates feature extraction and carries out dimensionality reduction as typical CNNs do. Although the implementation of an OSS accelerator with fewer nodes downgrades its classification accuracy, it is beneficial in terms of required minimal input power as fewer nodes translate to less splitting ratio. For instance, the exploitation of 4x4 patches on a 128 GSA/s modulating scheme dictates the use of 8 GHz PDs and a minimum of 100 μW per node without taking into consideration any other losses in the chip.

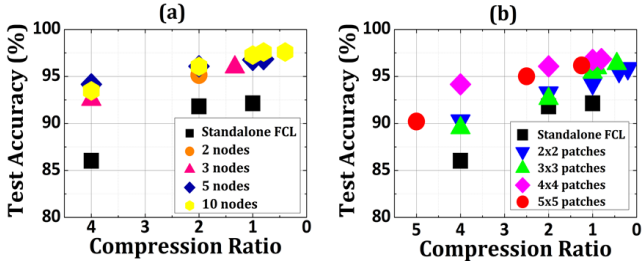

Fig. 3. Testing accuracy with respect to the compression ratio of the OSS-CNN: (a) on a 4x4 patching arrangement for a different number of nodes and (b) on a 5 node scheme for different patches.

The compute speed, power efficiency and compute density are calculated for the proposed accelerator following the methodology of [17,24]. Given that the MRR filter-nodes can be designed to impose negligible bandwidth limitation or latency, data processing rate of the OSS-CNN is only dictated by the rate of pixels fueling the convolutional stage. The pixel rate is therefore capped by the front-end DAC which can operate at a state-of-the art rate of 128 GSa/s [25] whereas optical modulation schemes can match this rate[26]. Furthermore, it should be noted that the proposed accelerator can also support wavelength division multiplexing (WDM) schemes, where information can be injected in parallel at wavelengths matching the free spectral range (FSR) of the filters. Therefore, the computation speed of the OSS-CNN can be expressed as:

$$CS = W \cdot n^2 \cdot N \cdot PR \quad (3)$$

where $W$ is the number of employed wavelengths, $n^2$ is the kernel/patch size and $N$ is the number of kernels which is equal to the number of OSS nodes. Assuming a 10-node OSS-CNN, a 128GSa/s pixel rate, a 4x4 kernel/patching scheme and a single wavelength the computational speed is approximately 20.5 TMAC/s or equivalently 41 TOPS. The total footprint corresponds to the total occupied area from the OSS nodes which is given by:

$$A = (2.2 \cdot 2 \cdot R) \cdot N \cdot (2.2 \cdot 2 \cdot R + \Delta h) \quad (4)$$

where $\Delta h$ is the distance between the OSS filter-nodes and $R$ is the ring radius of the MRR filters. Aiming to match the FSR of the MRR with the pixel rate, the ring radius is assumed to be 108 μm and a 10 μm distance between the nodes is set; the total footprint is equal to 2.32 mm². Therefore, the compute density of the OSS-CNN scales to 17.65 TOPS/mm². Moreover, the total power consumption is given in formula (5) according to [17] and in our case it is dominated by the electro-optic modulator and the ADCs.

$$P = N \left[ \frac{\frac{hv}{\eta} \max\left(2^{2N_b+1}, \frac{C_d V_r}{e}\right) PR}{n^2} + \frac{E_{mod} N_b PR}{N} + \frac{E_{ADC} N_b PR}{n^2} \right] W \quad (5)$$

where $\eta = \eta_L \cdot \eta_{MRR} \cdot \eta_{PD}$ is the combined total quantum efficiency of the detector ($\eta_{PD}$), laser ($\eta_L$) and optical loss through the MRR ($\eta_{MRR}$). Additionally, $hv$ is the energy of a single photon, $N_b$ the required bit-precision, $C_d$ the capacitance of the PD, $V_r$ the driving voltage of the PD, $E_{mod}$ the energy per bit of the modulator and $E_{ADC}$ the energy per bit of the ADC. It is assumed that $\eta_L = \eta_{PD} = 0.1$, whereas for $R = 108$ μm, 0.4 dB/cm losses and a coupling ratio equal to 0.1, the losses at the drop port are -3.5 dB corresponding to an efficiency of $\eta_{MRR} = 0.45$. Assuming a bit precision equal to $N_b = 5$, $E_{mod} = 1$ pJ/bit [27], $E_{ADC} = 2$ pJ/bit [27], 128 GHz modulation rate, 10 nodes/kernels, with 4x4 patches and a single wavelength, the total power consumption for a carrier at 1550 nm is approximately 1.42 W.

To demonstrate the merits of the OSS-CNN acceleration we compared it with two photonic and a digital state-of-the-art architecture. Inference accuracy for the MNIST task, clock speed, power efficiency and compute density are presented in Table 1.

Table 1. Accelerating Architecture, Clock Speed, Testing Accuracy, Power Efficiency and Computational Density

| Scheme | Clock (GHz) | MNIST Accuracy (%) | Power Efficiency (TOPS/W) | Compute Density (TOPS/mm²) |
|---|---|---|---|---|
| Nvidia Tesla P40 [28] | 1.3 | > 99 | 0.19 | 0.1 |
| DEAP[15] | 128 | 97.6 | 3.42 | - |
| Photonic tensor core [17] | 128 | 96.1 | 0.18 | 284 |
| **OSS-CNN** | 128 | **97.6** | **28.38** | **17.65** |

Furthermore, aiming to demonstrate the acceleration properties of the OSS concept, we fed the MNIST dataset to a sophisticated, yet parameter hungry, LeNet-5 CNN with a single FCL and a softmax layer at its front-end. The same dataset was fed to a simple passive 10 node photonic accelerator with an identical front-end. For this test accuracy and the number of FLOPS were computed and are presented in Table 2. It can be seen that a radical reduction in FLOPS by 98% is achieved with a marginal loss of accuracy of 1.3%.

Table 2. CNN Architecture, Number of FCLs at the Output, Testing Accuracy and Total Inference FLOPS

| Scheme | Number of FCLs (Layer Outputs) | Test Accuracy | FLOPS |
|---|---|---|---|
| LeNet-5 | 1 (10) | 98,9% | 736,000 |
| OSS-NN | 1 (10) | 97,6% | 14,600 |

We should highlight that the power efficiency as well as the compute density are derived for the photonic architectures considering the same modulation rate and the respective quantum efficiency for each system. It is evident that OSS-CNN provides superior performance in terms of power efficiency compared to all implementations by a factor in the order of x8, whereas at the same time offers similar accuracy. In particular, when compared to a similar photonic core, enhancement is preserved due to the fact that in our case the power-hungry convolutional operation is completely passive. In addition, the receiving end of the OSS accelerator operates at a rate which is $n^2$ times lower than the pixel rate, due to the averaging pooling; significantly reducing power consumption at the receivers. Finally, a more subtle difference of the proposed scheme compared to previous implementations is the fact that in our case, weight multiplication at the convolutional layer is not performed using real, trainable weights but on the contrary are complex and not strictly regulated, due to the fact that they originate from the shape of the transfer function of the employed filter. In this context, although the proposed scheme offers similar performance to schemes where all the CNN weights are trainable, it can be classified as an unconventional convolutional processor which performs convolution in the analog domain based on the linear and nonlinear properties of optoelectronic components similarly to what reservoir computing constitutes in the ecosystem of recurrent neural networks.

**Funding.** This work was funded by the EU H2020 NEoteRIC project (871330)

**Disclosures.** The authors declare no conflicts of interest.

**Data availability.** Data underlying the results presented in this paper are not publicly available at this time but may be obtained from the authors upon reasonable request.

## References

1. M. H. Miraz, M. Ali, P. S. Excell, and R. Picking, in *2015 Internet Technologies and Applications (ITA)* (2015), pp. 219–224.
2. J. Backus, Commun. ACM **21**, 613 (1978).
3. Y. Shen, N. C. Harris, S. Skirlo, M. Prabhu, T. Baehr-Jones, M. Hochberg, X. Sun, S. Zhao, H. Larochelle, D. Englund, and M. Soljačić, Nature Photonics **11**, 441 (2017).
4. G. Tanaka, T. Yamane, J. B. Héroux, R. Nakane, N. Kanazawa, S. Takeda, H. Numata, D. Nakano, and A. Hirose, Neural Networks **115**, 100 (2019).
5. D. Pérez, I. Gasulla, and J. Capmany, Opt. Express, OE **26**, 27265 (2018).
6. K. Sozos, A. Bogris, P. Bienstman, and C. Mesaritakis, in *2021 European Conference on Optical Communication (ECOC)* (2021), pp. 1–4.
7. A. Argyris, Nanophotonics **11**, 897 (2022).
8. J. Robertson, P. Kirkland, J. A. Alanis, M. Hejda, J. Bueno, G. Di Caterina, and A. Hurtado, Sci Rep **12**, 4874 (2022).
9. C. M. Valensise, I. Grecco, D. Pierangeli, and C. Conti, (2022).
10. K. Kitayama, M. Notomi, M. Naruse, K. Inoue, S. Kawakami, and A. Uchida, APL Photonics **4**, 090901 (2019).
11. A. R. Totovic, G. Dabos, N. Passalis, A. Tefas, and N. Pleros, IEEE J. Select. Topics Quantum Electron. **26**, 1 (2020).
12. J. Gu, Z. Wang, J. Kuen, L. Ma, A. Shahroudy, B. Shuai, T. Liu, X. Wang, G. Wang, J. Cai, and T. Chen, Pattern Recognition **77**, 354 (2018).
13. S. Chetlur, C. Woolley, P. Vandermersch, J. Cohen, J. Tran, B. Catanzaro, and E. Shelhamer, (2014).
14. M. Hu, J. P. Strachan, Z. Li, E. M. Grafals, N. Davila, C. Graves, S. Lam, N. Ge, J. J. Yang, and R. S. Williams, in *2016 53nd ACM/EDAC/IEEE Design Automation Conference (DAC)* (2016), pp. 1–6.
15. V. Bangari, B. A. Marquez, H. Miller, A. N. Tait, M. A. Nahmias, T. F. de Lima, H.-T. Peng, P. R. Prucnal, and B. J. Shastri, IEEE J. Select. Topics Quantum Electron. **26**, 1 (2020).
16. S. Xu, J. Wang, and W. Zou, IEEE Photon. Technol. Lett. **33**, 89 (2021).
17. J. Feldmann, N. Youngblood, M. Karpov, H. Gehring, X. Li, M. Stappers, M. Le Gallo, X. Fu, A. Lukashchuk, A. S. Raja, J. Liu, C. D. Wright, A. Sebastian, T. J. Kippenberg, W. H. P. Pernice, and H. Bhaskaran, Nature **589**, 52 (2021).
18. X. Xu, M. Tan, B. Corcoran, J. Wu, A. Boes, T. G. Nguyen, S. T. Chu, B. E. Little, D. G. Hicks, R. Morandotti, A. Mitchell, and D. J. Moss, Nature **589**, 44 (2021).
19. H. Bagherian, S. Skirlo, Y. Shen, H. Meng, V. Ceperic, and M. Soljacic, (2018).
20. S. M. Ranzini, R. Dischler, F. Da Ros, H. Bülow, and D. Zibar, Journal of Lightwave Technology **39**, 2460 (2021).
21. K. Sozos, A. Bogris, P. Bienstman, G. Sarantoglou, S. Deligiannidis, and C. Mesaritakis, Commun Eng **1**, 1 (2022).
22. "MNIST handwritten digit database, Yann LeCun, Corinna Cortes and Chris Burges," http://yann.lecun.com/exdb/mnist/.
23. S. Chen, X. Wang, C. Chen, Y. Lu, X. Zhang, and L. Wen, (2019).
24. M. A. Nahmias, T. F. de Lima, A. N. Tait, H.-T. Peng, B. J. Shastri, and P. R. Prucnal, IEEE J. Select. Topics Quantum Electron. **26**, 1 (2020).
25. F. Buchali, V. Aref, R. Dischler, M. Chagnon, K. Schuh, H. Hettrich, A. Bielik, L. Altenhain, M. Guntermann, R. Schmid, and M. Moller, J. Lightwave Technol. **39**, 763 (2021).
26. M. Burla, C. Hoessbacher, W. Heni, C. Haffner, Y. Fedoryshyn, D. Werner, T. Watanabe, H. Massler, D. Elder, L. Dalton, and J. Leuthold, in *2019 Conference on Lasers and Electro-Optics (CLEO)* (2019), pp. 1–2.
27. M. A. Al-Qadasi, L. Chrostowski, B. J. Shastri, and S. Shekhar, APL Photonics **7**, 020902 (2022).
28. "Inference Platforms for HPC Data Centers from NVIDIA Deep Learning AI," https://www.nvidia.com/en-au/deep-learning-ai/inference-platform/hpc/.